\documentclass{aa} 
\usepackage{graphicx}
\usepackage{txfonts}
\begin{document}
   \title{Shape model and spin state of non-principal axis rotator (5247)~Krylov}

   \subtitle{}

   \author{H.-J. Lee
          \inst{1}\fnmsep\inst{2}\and
          J. \v{D}urech\inst{3}\and
          M.-J. Kim\inst{2}\and
          H.-K. Moon\inst{2}\and
          C.-H. Kim\inst{1}\and
          Y.-J. Choi\inst{2}\fnmsep\inst{4}\and
          A. Gal\'ad\inst{5}\and
          D. Pray\inst{6}\and
          A. Marciniak\inst{7}\and
          M. Kaplan\inst{8}\and
          O. Erece\inst{8}\fnmsep\inst{9}\and
          R. Duffard\inst{10}\and
          \v{S}. Gajdo\v{s}\inst{5}\and
          J. Vil\'agi\inst{5}\and
          M. Lehk\'y\inst{3}
          }

   \institute{Chungbuk National University, 1 Chungdae-ro, Seowon-gu, Cheongju, Chungbuk 28644, Korea\\
              \email{hjlee@kasi.re.kr}
         \and
             Korea Astronomy and Space Science Institute, 776, Daedeokdae-ro, Yuseong-gu, Daejeon 34055, Korea\and
             Astronomical Institute, Faculty of Mathematics and Physics, Charles University, V Hole\v{s}ovi\v{c}k\'ach~2, 180 00 Prague~8, Czech Republic\and
             University of Science and Technology, 217, Gajeong-ro, Yuseong-gu, Daejeon 34113, Korea\and
             Modra Observatory, Department of Astronomy, Physics of the Earth, and Meteorology, FMPI UK, Bratislava SK-84248, Slovakia\and
             Sugarloaf Mountain Observatory, South Deerfield, MA, USA\and
             Astronomical Observatory Institute, Faculty of Physics, Adam Mickiewicz University, S{\l}oneczna 36, 60-286 Pozna\`n, Poland\and
             Akdeniz University, Department of Space Sciences and Technologies, 07058 Antalya, Turkey\and
             T\"{U}B\.ITAK National Observatory, Akdeniz University Campus, 07058 Antalya, Turkey\and
             Departamento de Sistema Solar, Instituto de Astrof{\'i}sica de Andaluc{\'i}a (CSIC), Glorieta de la Astronom{\'i}a s/n, 18008 Granada, Spain
             }

   \date{Received Sep 30 2019; accepted Feb 05 2020}


  \abstract
   {The study of non-principal axis (NPA) rotators can provide important clues to the evolution of the spin state of asteroids. However, so far, very few studies have focused on NPA-rotating main-belt asteroids (MBAs). One of MBAs that are known to be in an excited rotation state is asteroid (5247)~Krylov.}
   {By using disk-integrated photometric data, we construct a physical model of (5247) Krylov including shape and spin state.}
   {We apply the light curve convex inversion method employing optical light curves obtained by using ground-based telescopes in three apparitions during 2006, 2016, and 2017, along with infrared light curves obtained by the Wide-field Infrared Survey Explorer (WISE) satellite in 2010.}
   {Asteroid (5247)~Krylov is spinning in a short axis mode (SAM) characterized by rotation and precession periods of 368.7 hr and 67.27 hr, respectively. The angular momentum vector orientation of Krylov is found to be $\lambda_{L} = 298^\circ$ and $\beta_{L} = -58^\circ$. The ratio of the rotational kinetic energy to the basic spin state energy $E/E_{0} \simeq 1.02$ shows that the (5247)~Krylov is about 2\% excited state compared to the Principal Axis (PA) rotation state. The shape of (5247)~Krylov can be approximated by an elongated prolate ellipsoid with a ratio of  moments of inertia of $I_{a}:I_{b}:I_{c}=0.36:0.96:1$. This is the first physical model of NPA rotator among MBAs. The physical processes that led to the current NPA rotation cannot be unambiguously reconstructed.}
   {}
   \keywords{Minor planets, asteroids: individual -- Techniques: photometric}
   \maketitle

\section{Introduction}
\label{section:introduction}
    The spin state and shape are basic physical properties of asteroids. They are related to orbital evolution due to the thermal Yarkovsky effect \citep{Vokrouhlicky_et_al_2015}, collisional evolution \citep{Bottke_et_al_2015}, and rotational disruption \citep{Walsh_et_al_2015}.

    The spin state of an asteroid can be classified as principal axis (PA) rotation or non-principal axis (NPA) rotation. The PA rotation refers to the state in which the energy for a given angular momentum is minimized and the angular momentum vector is aligned with the axis of rotation. The NPA rotation state refers to an excited spin state in which the angular momentum vector is misaligned with the rotation axis. This type of rotation has also been referred to as ``tumbling'' by \citet{Harris_1994}, since then, it is commonly used when referring to NPA rotation.

    Investigations of tumbling asteroids have attracted continuous attention since the discovery of the NPA rotation of (4179)~Toutatis by radar observations of \citet{Hudson_et_al_1995}. In particular, several previous studies have focused on processes to understand the spin evolution of NPA rotating asteroids via simulation analyses. Spin evolution processes were summarized based on previous studies by \citet{Pravec_et_al_2014}. According to them, the tumbling motion of asteroids can occur because of the following four reasons: 1) original tumbling, 2) sub-catastrophic impact, 3) spin down by the YORP effect, and 4) effect of gravitational torque during planetary flyby. Additionally, many studies have been performed on tumbling asteroids that evolve into PA rotators due to rotational kinetic-energy dissipation via cyclic variations in stresses and strains (\citealt{Prendergast_1958}; \citealt{Pravec_et_al_2014}, reference therein).

    Constructing the spin state of individual tumblers may in some cases restrain the most likely stimulation mechanism, and having a statistically significant sample of models of NPA rotators would help us to match the observed population of tumblers with the theoretically described mechanisms of rotation excitation and damping.

    The spin state of an NPA rotator can be revealed by radar observations or time-series disk-integrated photometry. However, spin-state analysis via radar observations can be restrictively conducted only on close-approaching objects or a few large asteroids. In addition, even with radar observations, it is difficult to observe asteroids for sufficiently long periods of time. Therefore, NPA rotation analysis based on time-series photometry has attracted extensive attention. In particular, a method to analyze an NPA rotator using light curves was first proposed by \citet{Kaasalainen_2001}. Thereafter, \citet{Pravec_et_al_2005} conducted a period analysis of NPA rotators and studied their spin states assuming that the shape of the asteroid is a simple triaxial ellipsoid. Thus far, physical models of only four NPA rotators (all of them are near-Earth asteroids) have been constructed: 2008~TC3 \citep{Scheirich_et_al_2010}, (214869)~2007~PA8 \citep{Brozovic_et_al_2017}, (99942)~Apophis \citep{Pravec_et_al_2014}, and (4179)~Toutatis \citep{Hudson_et_al_1995}. Of these, models of (214869)~2007~PA8 and (4179)~Toutatis were obtained based on radar observations, whereas spin states and shape models of the other two asteroids were determined based on light-curve analysis. Very recently, the interstellar object `Oumuamua was found to have NPA rotation \citep{Belton_et_al_2018, Drahus_et_al_2018, Fraser_et_al_2018}, and later \citet{Mashchenko_2019} attempted to construct its shape and spin state using a physical model.

    So far, the construction of shape and NPA rotation model has been conducted on the near-Earth asteroids (NEAs) or an interstellar object, and no main-belt asteroids (MBAs). However, the number of NPA rotators known to date comprises 69 MBAs and 111 NEAs based on LCDB (Lightcuve Database; version August 2019; \citealt{Warner_et_al_2009}). Because the MB is relatively stable compared to other regions on the timescale of the solar system, MBAs are advantageous for studying the history of asteroids. Therefore, knowing the spin state and shape model of NPA rotators in the MB is useful for understanding spin-state evolution mechanism of NPA rotators.

    Hence, to study the spin states of NPA rotators existing in the MB, we analyzed the spin state of (5247) Krylov (1982 UP6) (hereafter Krylov). Krylov was discovered by Karachkina on 20 Oct. 1982 in the Nauchnyj observatory \citep{Marsden_Williams_1993}. Based on its orbital characteristics, Krylov has been classified with the hierarchical clustering method as belonging to the Phocaea collisional family \citep{Nesvorny_2015}. NPA rotation of Krylov was first reported by \citet{Pravec_et_al_2006}. Following this, \citet{Lee_et_al_2017} confirmed NPA rotation of Krylov based on its double-period light curve ($P_1 = 82.188$\,hr, $P_2 = 67.13$\,hr), and they classified the taxonomy of this asteroid as the S-type. In addition, the diameter of Krylov has been estimated independently by infrared observations of satellites AKARI and NEOWISE: however, the AKARI diameter ($10.44 \pm 0.37$\,km; \citealt{Usui_et_al_2011}) does not match well with the NEOWISE diameters ($7.716 \pm 0.043$ or $8.665 \pm 0.557$\,km; \citealt{Mainzer_et_al_2019}).

    As mentioned above, Krylov is an NPA rotating MBA observed at various wavelengths. However, no attempt to determine a shape model of Krylov was made so far. In this paper, based on the historical photometric data and new disk-integrated photometry, we present the first shape model as well as the improved spin state of Krylov.

    This paper is organized as follows: In Sect.~\ref{section:disk-integrated_photometry}, we describe light curve data used for the convex light curve inversion method (\citealt{Kaasalainen_2001}; \citealt{Kaasalainen_Torppa_2001}; \citealt{Kaasalainen_et_al_2001}). A physical model of Krylov is presented in Sect.~\ref{section:physical_model}. In Sect.~\ref{section:discussion}, we discuss its spin state and the possible evolutionary process.

%

   \begin{table*}
      \caption{Observational and instrument details}
         \label{table:1}
            \centering
            \begin{tabular}{l l l l l }     
            \hline\hline
            Observatory	& Telescope aperture &	Duration	&	Detector (Filter)	&	Ref.\\
            \hline
            Modra                        &  0.6 m & Jun. -- Jul. 2006 & AP8p (clear)          & This work\\
                                         &        & (21 nights)    &                &          \\
            SMO                          & 0.35 m & May. -- Jun. 2006  & SBIG ST-10XME (clear)  & This work\\
                                         &        & (12 nights)    &                &          \\
            WISE                         & 0.4 m  & Jun. 2010 & Teledyne HgCdTe (W1 \& W2) & \citet{Wright_et_al_2010}\\
                                         &        & (1 night)    & DRS Si:As (W3 \& W4) & \citet{Mainzer_et_al_2011} \\
            KMTNet                       & 1.6 m  & Jan. -- Apr. 2016 & 18K mosaic CCD (R) & \citet{Lee_et_al_2017}\\
            (CTIO, SAAO, and SSO)        &        & (51 nights)    &with four e2v 9K&          \\
            TUG                          & 1.0 m  & Jul. -- Aug. 2017  & SI 4K (R)          & This work\\
                                         &        & (4 nights)     &                &          \\
            LOAO                         & 1.0 m  & Jun. -- Sep. 2017  & e2v 4K (R)         & This work\\
                                         &        & (4 nights)     &                &          \\
            BOAO                         & 1.8 m  & Jun. -- Sep. 2017  & e2v 4K (R)         & This work\\
                                         &        & (6 nights)     &                &          \\
            OAdM                         & 0.8 m  & Jun. -- Aug. 2017  & e2v 2K (R)         & This work\\
                                         &        & (7 nights)     &                &          \\
            La Sagra                     & 0.45 m & Jun. 2017      & SBIG ST-10XME (R)  & This work\\
                                         &        & (3 nights)     &                &          \\
            McDonald                     & 2.1 m  & Jul. 2017      & SQUEAN (r)         & This work\\
                                         &        & (6 nights)     &                &          \\
            BlueEye 600                  & 0.6 m  & Jul. 2017      & G4-4000BI (R)      & This work\\
                                         &        & (2 nights)     &                &          \\
            \hline
            \end{tabular}
            \tablefoot{SMO = Sugarloaf Mountain Observatory, WISE = Wide-field Infrared Survey Explorer, KMTNet = Korea Microlensing Telescope Network, CTIO = Cerro Tololo Inter-American Observatory, SAAO = South African Astronomical Observatory, SSO = Siding Spring Observatory, TUG = T\"{U}B\.ITAK National Observatory, LOAO = Lemmonsan Optical Astronomy Observatory, BOAO = Bohyunsan Optical Astronomy Observatory, OAdM = Montsec Astronomical Observatory.}
   \end{table*}
   
\section{Disk-integrated photometry}
\label{section:disk-integrated_photometry}

    We constructed a physical model of Krylov using ground-based optical light curves and the infrared light curve observed from the WISE satellite. In order to obtain a rotational light curve of Krylov, we took optical imaging data on a total of 116 nights in 2006, 2016, and 2017. The dataset collected over 51 nights in 2016 at the KMTNet three sites \citep{Kim_et_al_2016}, was already published by \citet{Lee_et_al_2017}, whereas the data from 2006 and 2017 have not been published before. All the observations were made using multiple 0.35--2.1-m telescopes equipped with charge-coupled device (CCD) cameras. Moreover, we gathered infrared fluxes of Krylov from the WISE catalog (\citealt{Wright_et_al_2010}; \citealt{Mainzer_et_al_2011}). The detailed observation information is provided in Table~\ref{table:1}. The geometries and observational circumstances are listed in Appendix A.

    In the optical observations, we determined the exposure time by considering the brightness and sky motion of Krylov and the seeing condition during observation ensuring that the asteroid remained as a point source. In addition, our analysis did not include contaminated images where the asteroid overlaps with field stars.

\begin{figure*}[t!]
   \centering
    \includegraphics[width=140mm]{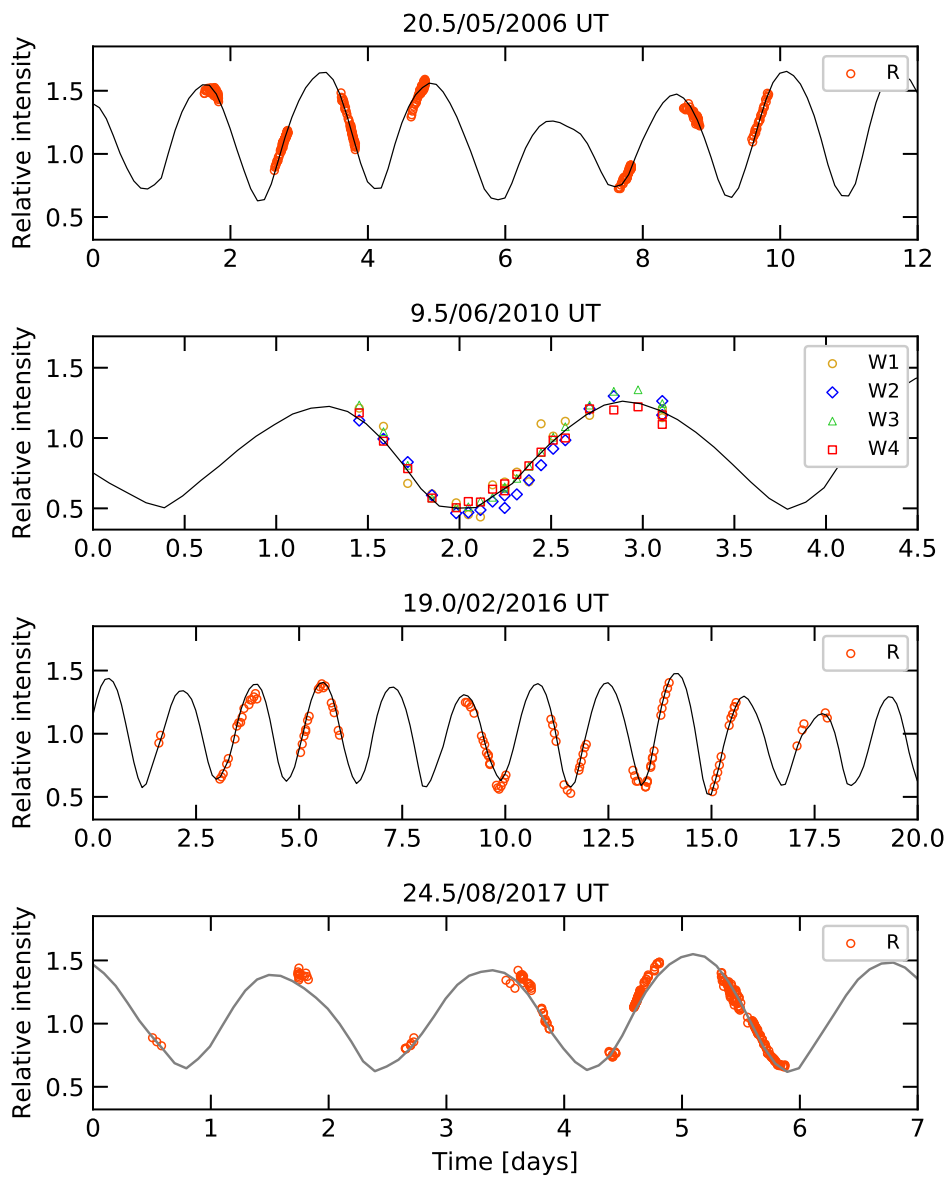}
      \caption{Example light curves of Krylov from four different apparitions; 2006 : this work, 2010 : \citet{Wright_et_al_2010}; \citet{Mainzer_et_al_2011}, 2016 : \citet{Lee_et_al_2017}, 2017 : this work. Solid curves represent synthetic light curves provided by the best-fit solution. The start times of each light curve are mentioned in the titles.}
         \label{Fig:lightcurves}
   \end{figure*}

    The 2006 observation data were reduced by the pipeline of each observatory. Preprocessing of the raw frames from Sugarloaf Mountain Observatory (SMO) was conducted using the Maxim DL program. In this process, bias, dark, and flat-field corrections were applied. By using Canopus software \citep{Warner_2006}, photometry was carried out to obtain instrumental magnitudes of stars in the frames. Similarly, the raw frames observed at Modra Observatory were processed with the same Maxim DL program for the bias, dark, and flat-field corrections. Aperture photometry was used to obtain instrumental magnitudes of the frames. Both sets of light curves observed at SMO and Modra were created using differential photometry. Since different comparison stars were used on different nights, these light curves show to be fragmented. Nonetheless we did not calibrate the relative differences between the light curves. Instead, the differences were adjusted using the inversion method based on relative brightness which was developed by \citet{Kaasalainen_Torppa_2001}. More details of the inversion method will be discussed in Sect.~\ref{section:physical_model}. A sample of the 2006 light curves merged by the inversion method is shown in the top panel of Fig.~\ref{Fig:lightcurves}.

    Using the Image Reduction and Analysis Facility (IRAF) software package, the photometric data from the apparition in 2017 were reduced in a consistent way to obtain a calibrated light curve from data collected at various stations. Preprocessing was carried out using the IRAF/CCDRED package. We corrected the bias, dark, and flat-field images during preprocessing. Further, we calculated the World Coordinate System (WCS) solution via matching with the USNO B1.0 catalog upon employing the SCAMP package \citep{Bertin_2010}. Aperture photometry of these images was carried out using the IRAF/APPHOT package. We set the aperture radius to the full-width half maximum of the stellar profile to obtain the maximum signal to noise ratio \citep{Howell_1989}. The standard magnitudes were determined and applied to the ensemble normalization technique \citep{Gilliland_et_al_1988, Kim_et_al_1999} with the Pan-STARRS Data Release 1 catalog \citep[PS DR1;][]{Chambers_et_al_2016}. The magnitudes of PS DR1 were converted to Johnson--Cousins filter magnitude based on empirical transformation equations \citep{Tonry_et_al_2012}. Thus, all of the light curves observed in 2017 were obtained as absolute light curves. The bottom panel of Fig.~\ref{Fig:lightcurves} shows a part of the 2017 light curves.
    In addition, a part of the 2016 light curves by \citet{Lee_et_al_2017} was reproduced in the third panel of Fig.~\ref{Fig:lightcurves}.

    Thermal light curves are useful for constructing physical models of asteroids when analyzed in conjunction with optical light curves \citep{Durech_et_al_2018}. Furthermore, as the observation period of the WISE light curve corresponds with the optical light curve “gap” between 2006 and 2016, this dataset can assist in a more precise analysis of the physical model. Therefore, we included infrared light curve in our analysis. Infrared data was obtained on June 11, 2010 UT in four infrared bands at 3.4, 4.6, 12, and 22 $\mu$m, usually referred to as W1, W2, W3, and W4, respectively. In addition, we used only the measurements with quality flags A, B, or C, and artifact flags 0, p, or P except for data flagged as potentially affected by artifact contamination as per the WISE moving object pipeline subsystem (\citealt{Cutri_et_al_2012}). We assumed that the shape of thermal light curves is similar to that of reflected light and treated them as relative light curves in the same way as \citet{Durech_et_al_2018}. The second panel of Fig.~\ref{Fig:lightcurves} shows the WISE light curve.

%

   \begin{figure*}[t]
   \centering
    \includegraphics[width=140mm]{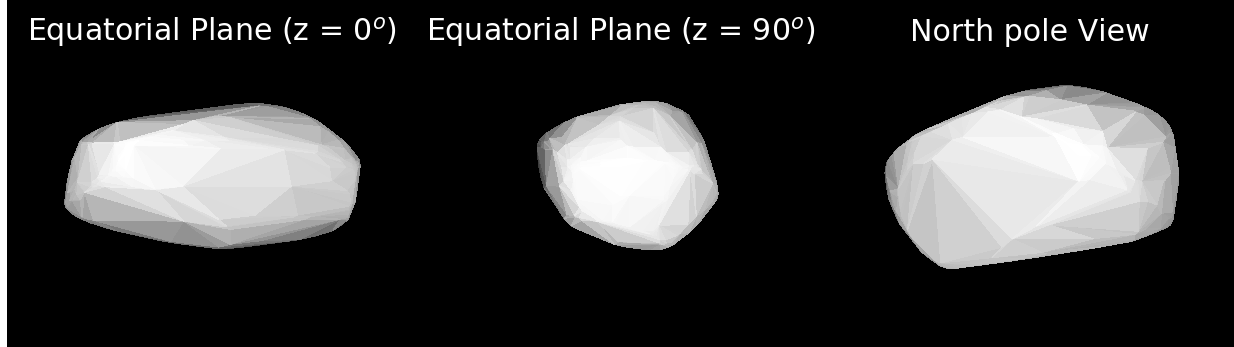}
      \caption{Convex shape model of Krylov.}
         \label{Fig:shape}
   \end{figure*}
%

\section{Physical model}
\label{section:physical_model}
    Simultaneous analysis of light curves was performed by application of the light curve convex inversion method for NPA rotators described by \citet{Kaasalainen_2001}, \citet{Kaasalainen_Torppa_2001}, \citet{Kaasalainen_et_al_2001}. The code for NPA light curve inversion was first developed in Fortran by M. Kaasalainen and later converted to C language by co-author J. \v{D}urech. The spin state of an NPA rotator can be represented by eight parameters ($\lambda_{L}$, $\beta_{L}$, $\phi_{0}$, $\psi_{0}$, $P_{\psi}$, $P_{\phi}$, $I_{a}$, $I_{b}$; \citealt{Kaasalainen_2001, Pravec_et_al_2005, Scheirich_et_al_2010, Pravec_et_al_2014}). Parameters $\lambda_{L}$ and $\beta_{L}$ denote the ecliptic longitude and ecliptic latitude, respectively, of the constant angular momentum vector $\mathbf{L}$ of the NPA rotator. $\phi_{0}$ and $\psi_{0}$ are standard Euler angles at $t_{0}$. These Euler angles are defined as angles between principal axes of asteroids and the inertial coordinate system. The inertial frame of the Z-axis is aligned with the angular momentum vector, and the XZ plane includes a vector pointing to the vernal equinox. The third Euler angle, $\theta_{0}$, is not used as an independent parameter because it can be calculated from other parameters. Angles $\phi$, $\theta$, and $\psi$ refer to the angle of precession, nutation (or tilt), and rotation, respectively. Further, $P_{\psi}$ and $P_{\phi}$ indicate the rotation and precession periods, respectively. $I_{a}$ and $I_{b}$ are the moments of inertia corresponding to the principal axes of the asteroid body. These moments of inertia are normalized by the moment of inertia along the rotation axis, $I_{c}$. In addition, NPA rotation can be classified into two modes: short axis mode (SAM) and long axis mode (LAM) according to the rotation axis. SAM refers to rotation around the shortest axis while LAM involves rotation around the longest axis.
    
    As the light curves obtained from observations in 2006 and 2010 were not absolute--calibrated, we analyzed data from apparitions in 2016 and 2017 independently prior to analyzing the whole data set. Before modeling each light curves was conducted, we determined $P_{\psi}$ and $P_{\phi}$. According to period analysis of the simulated light curve of a tumbling asteroid, the primary ($f_{1}$) and secondary ($f_{2}$) frequencies are usually determined by combining the rotation and precession frequencies, $2f_{\phi}$, 2($f_{\phi}  \pm f_{\psi}$) (where $+$ sign is for LAM and $-$ for SAM), and $f_{\phi}$ \citep{Kaasalainen_2001}. Thus, we checked the possible frequency combinations based on the results of a previous study by \citet{Lee_et_al_2017}, with $f_{1} = $ 0.58402 cycles/day and $f_{2} = 0.7124$ cycles/day. As a result, we selected three candidates for frequency combination, $f_1  = 2 f_\phi$ and $f_2 = 2 (f_\phi + f_\psi)$ (LAM); $f_1 = 2(f_\phi - f_\psi)$ and $f_2 = 2 f_\phi$ (SAM1); $f_1 = f_\phi$ and $f_2  = 2(f_\phi - f_\psi)$ (SAM2).
    
    We tested the optimization of the physical model for each of these candidates. For each candidate, we constructed a grid of parameters ($\lambda_{L}$, $\beta_{L}$, $\phi_{0}$, $\psi_{0}$, $I_{a}$, $I_{b}$). The grid for the orientation of the angular momentum vector was constructed by distributing ten different positions evenly on the celestial sphere. Standard Euler angles at $t_{0}$ were arranged at 60$^\circ$ intervals. In addition, the moments of inertia were sequenced at 0.01 intervals from 0.01 to 0.99 in the case of SAM, and intervals of 0.1 from 1.1 to 10 in the case of LAM. Optimization was performed using the grid as the initial parameter set. The photometric model of the surface properties of airless body was considered as the Hapke model \citep{Hapke_1993}. The Hapke model parameters were optimized using initial values of a typical S-type asteroid: $\varpi = 0.23$, $g = -0.27$, $h = 0.08$, $B_{0} = 1.6$, and $\bar{\theta} = 20^\circ$ \citep{Li_et_al_2015}, where $\varpi$ denotes the single scattering albedo, $g$ denotes the particle phase function parameter, $h$ and $B_{0}$ denote the width and amplitude of the opposition surge, respectively, and $\bar{\theta}$ denotes the macroscopic roughness angle. Additionally, because light curves used in this study covered only phases angles from 14.0$^\circ$ to 28.5$^\circ$, the parameters for opposition surge ($h$ and $B_{0}$) were fixed during the optimization process. The $\bar{\theta}$ was also fixed because it cannot be reliably obtained from our data. Thus, two parameters only, $\varpi$ and $g$, were optimized. We found that the physical model is not sensitive to the Hapke model parameters. Solutions corresponding to the 2016 and 2017 data converged at one or two global minima for each frequency combination. However, we found that the values between the moment of inertia obtained from the shape model (assuming constant density) and the dynamical moments of inertia were different in all solutions except for SAM1. Because we preferred physically self-consistent model, we excluded the combination of LAM and SAM2. In addition, solutions corresponding to the 2016 and 2017 data were similar to SAM1. Therefore, we accepted values for $P_{\psi}$ and $P_{\phi}$ based on SAM1.

   \begin{table}
      \caption{Physical model of Krylov}
         \label{table:2}
            \centering
            \begin{tabular}{l l}     
            \hline\hline
            Physical parameter	                                 & Value\\
            \hline
            Fitted parameters                                    & \\
            $\lambda_{L}$ [deg]                                  & $298 \pm 37$\\
            $\beta_{L}$ [deg]                                    & $-58 \pm 12$\\
            $P_{\psi}$ [hr]                                      & $368.7 \pm 0.2$\\
            $P_{\phi}$ [hr]                                      & $67.27 \pm 0.01$\\
            $\psi_{0}$ [deg]                                     & $5 \pm 28$\\
            $\phi_{0}$ [deg]                                     & $126 \pm 14$\\
            $t_{0}$ [day]                                        & 2453877.618993\\
            $I_{a}/I_{c}$                                        & $0.36 \pm 0.02$\\
            $I_{b}/I_{c}$                                        & $0.96 \pm 0.01$\\
            Derived parameters                                   & \\
            $(P_{\phi}^{-1} - P_{\psi}^{-1})^{-1} = P_{1}$ [hr]  & $82.28 \pm 0.04$\\
            $\theta_\mathrm{aver}$ [deg]                                & $30 \pm 10$\\
            $\theta_\mathrm{min}$ [deg]                                 & $6 \pm 5$\\
            $\theta_\mathrm{max}$ [deg]                                 & $45 \pm 20$\\
            $a_\mathrm{dyn}/c_\mathrm{dyn}$                                    & $2.2 \pm 0.1$\\
            $b_\mathrm{dyn}/c_\mathrm{dyn}$                                    & $1.12 \pm 0.02$\\
            $a_\mathrm{shape}/c_\mathrm{shape}$                                & $2.1 \pm 0.4$\\
            $b_\mathrm{shape}/c_\mathrm{shape}$                                & $1.13 \pm 0.2$\\
            $E/E_0$                                              & $1.023 \pm 0.001$\\
            \hline
            \end{tabular}
            \tablefoot{$\theta_\mathrm{aver}$, $\theta_\mathrm{min}$, and $\theta_\mathrm{max}$ : the average, minimum, and maximum value of $\theta$ over one cycle; $a_\mathrm{dyn}/c_\mathrm{dyn}$ and $b_\mathrm{dyn}/c_\mathrm{dyn}$ : the axial ratio of a dynamically equivalent ellipsoid; $a_\mathrm{shape}/c_\mathrm{shape}$ and $c_\mathrm{shape}/c_\mathrm{shape}$ : the axial ratio of a convex shape model; $E/E_0$ : the ratio of the rotational kinetic energy to the lowest energy for the given angular momentum.}
   \end{table}
   
    The final physical model of Krylov was obtained by optimizing the entire set of light curves in 2006, 2010, 2016, and 2017. Since the 2006 and 2010 light curves were obtained from differential photometry rather than absolute one, as discussed in the previous section, they were analyzed using an inversion method based on relative brightness \citep{Kaasalainen_Torppa_2001}. In the inversion method, both the observed and modeled light curves of each night are renormalized to mean brightness of unity before optimization is conducted. Because the light curves of 2016 exhibited the best phase coverage among all observation data, we used the solution of this light curve as the initial parameter set in our final optimization. Uncertainties in physical parameters were estimated from the 3 $\sigma$ interval of solutions from the light curve inversion method using a thousand bootstrapped \citep{press_et_al_1986} photometric datasets. Our final solution is presented in Table~\ref{table:2} and the convex shape model for this solution is presented in Fig.~\ref{Fig:shape}. The synthetic light curves of the best solution with real data are shown in Fig.~\ref{Fig:lightcurves}. The best-fit solution is very similar to the solution obtained without the inclusion of the WISE light curve. The infrared light curve was only used to confirm our final solution.

    The physical model of Krylov in Table 2 is summarized as follows: (1) Krylov is rotating in a SAM state with rotation and precession periods of 368.7 hr and 67.27 hr, respectively. (2) The angular momentum vector orientation of Krylov is located at ($298^\circ, -58^\circ$) in ecliptic coordinates. (3) The dynamical shape of Krylov looks like an elongated prolate ellipsoid with a ratio of  moments of inertia of $I_{a}:I_{b}:I_{c}=0.36:0.96:1$, as produced in Fig.~\ref{Fig:shape}. (4) The ratio of the rotational kinetic energy to the basic spin state energy $E/E_{0} \simeq 1.02$ indicates that Krylov is a NPA rotator with only 2\% excited state relative to the PA rotation state.

   \begin{table*}
      \caption{Parameters of AKARI, WISE1, and WISE2 observations.}
         \label{table:3}
            \centering
            \begin{tabular}{c c c c c c c c c c}     
            \hline\hline
            	& Date UT &	Band &  \# of points  &	Band &  \# of points  &	Band &  \# of points  &	Band &  \# of points\\
            \hline
            AKARI                        &  Sep. 15, 2006 UT & 18 $\mu$m & 3  & & & & & &\\
                                         &  Aug. 19, 2007 UT & 18 $\mu$m & 1  & & & & & &\\
            WISE1                        &  Jun. 11, 2010 UT & W1 & 18  & W2 & 19 & W3 & 20 & W4 & 20\\
            WISE2                        &  Dec. 03, 2010 UT & W1 & 7   & W2 & 7 & & & &\\
            \hline
            \end{tabular}
   \end{table*}

\section{Discussion}
\label{section:discussion}
    In this section, we examined some physical effects acting on Krylov to understand its current spin state and rotational evolution. Since Krylov is an MBA and belongs to the Phocaea asteroid family with age of about 2.2 Gyr \citep{Carrube_2009}, we considered three possible mechanisms; YORP effect, damping effect, and sub-catastrophic impact. According to the physical model of Krylov in Table 2, its excitation level is in about 2\% excited state compared to a PA rotation state. A comparison of the excitation level with those of other known tumblers with available physical models shows that the excitation level of Krylov is similar to those of (99942)~Apophis ($E/E_{0} \sim 1.02$; \citealt{Pravec_et_al_2014}) and (214869)~2007~PA8 ($\sim$ 1.08; \citealt{Brozovic_et_al_2017}), and quite lower than those of 2008~TC3 ($\sim$ 1.21; \citealt{Scheirich_et_al_2010}),  `Oumuamua ($\sim$ 1.98; \citealt{Mashchenko_2019}), and (4179)~Toutatis ($\sim$ 2.08; \citealt{Hudson_et_al_1995}). The small difference in the spin state of Krylov compared to its PA-rotation state can be attributed to (1) excitation of the asteroid by a low-magnitude force and (2) substantial damping of its NPA spin state from a higher excited-rotation state.

    In order to find the most likely mechanism for maintaining its current spin state, we estimated timescales for Krylov. The physical parameters of Krylov used to estimate the timescales are as follows: $D = 7.716$\,km  \citep{Mainzer_et_al_2019}, $a = 2.33$\,AU (Minor Planet Center), and $P_{1} = 82.28$\,hr (This work). $P_{1}$ is the strongest apparent period, the conjunction period between $P_{\psi}$ and $P_{\phi}$. The diameter is one of the parameters to calculate the time scale. However, Krylov diameters determined in previous studies are inconsistent with each other : 10.44 $\pm$ 0.37 km (AKARI; \citealt{Usui_et_al_2011}), 7.716 $\pm$ 0.043 km (WISE1; \citealt{Mainzer_et_al_2019}), and 8.655 $\pm$ 0.557 km (WISE2; \citealt{Mainzer_et_al_2019}). These discrepancies are not critical however because the timescales are based on order of magnitude estimate. Nevertheless, to clarify its diameter, we examined observation data used in each of these studies (refer to Table~\ref{table:3}). AKARI and WISE2 data were acquired using a lower number of observational passbands and fewer observation points than WISE1. Therefore, the diameter based on WISE1 data was considered the most reliable, and was used for timescale estimation.

    The collisional timescale ($\tau_\mathrm{collision}$) for an MBA can be estimated as:
    $$
    \tau_\mathrm{collision} = 16.8\,\mathrm{Myr} \sqrt{R} \simeq 1044\,\mathrm{Myr}\,,
    $$
    where $R$ is the radius of an asteroid in meters \citep{Farinella_et_al_1998}.

    At the same time, we considered the damping timescale ($\tau_\mathrm{damping}$) of a tumbling asteroid. It is the timescale over which NPA rotation shifts to PA rotation due to the dissipation of stress and strain forces. This timescale is calculated as:
    $$
    \tau_\mathrm{damping} = \frac{P_{1}^{3}}{C^{3} D^{2}} \simeq 200\,\mathrm{Myr}\,,
    $$
    where $C$ denotes a constant of 36, and $D$, $P_{1}$, and $\tau_\mathrm{damping}$ are in kilometers, hours, and Gyr \citep{Pravec_et_al_2014}, repectively.

    In addition, we estimated the YORP timescale ($\tau_\mathrm{YORP}$) for reaching the onset of tumbling spin-state driven by the YORP effect as:
    $$
    \tau_{YORP} = \tau_{0} \left(\frac{D}{D_{0}}\right)^{2} \left(\frac{a}{a_{0}}\right)^{2} \simeq 170\,\mathrm{Myr}\,,
    $$
    where $\tau_{0} = 5.3\,\mathrm{Kyr}$ was the previously determined value for $D_{0} = 50$\,m, $a_{0} = 2$\,AU and a reference initial rotation period $P_{0}=6$\,hr as reported by \citet{Vokrouhlicky_et_al_2007}. Should the initial rotation period be longer than $P_{0}$, the estimated YORP timescale would be shorter. On the other hand, the above-mentioned result was based on a simple approach to YORP effect considering only large scale shape modeling. Small scale irregularities of an asteroid surface, unresolved by our model, can cause typically diminish the strength of the resulting YORP effect. A similar trend has also been noted when comparing detected YORP values with their predictions \citep{Vokrouhlicky_et_al_2015}. A fudge factor of 2 to 3 in extending the estimated YORP timescale may be thus expected.

    From the timescales calculated above, it is very difficult to understand why Krylov becomes an NPA spinner with an excitation level of about 2\%. Nonetheless, it seems certain for Krylov that  the timescale of the YORP effect is comparable to that of the damping effect. This may imply  that the rotational kinetic energy loss by the strain-stress force may be simultaneously balanced by the YORP effect. Further, it may be supposed that occasional collisions within the age of the Phocaea asteroid family maintain Krylov's present NPA rotation against the PA rotation.

\begin{acknowledgements}

      We are very thankful to the reviewer, Dr. M. Drahus for his positive criticisms, suggestions, and comments which greatly improved the original version of the manuscript. This research is supported by Korea Astronomy and Space Science Institute (KASI). Work at KASI was partly supported under the framework of international cooperation program managed by the National Research Foundation of Korea (2017K2A9A1A06037218, FY2018). The work of J.~\v{D}urech and M.~Lehk\'y was supported by the grant 18-04514J of the Czech Science Foundation. The work of C.-H. Kim was financially supported by the Research Year of Chungbuk National University in 2018. The work at Modra was supported by the Slovak Grant Agency for Science VEGA, Grant 1/0911/17. Murat Kaplan and Orhan Erece thank to T\"{U}B\.ITAK for a partial support in using T100 telescope with project number 14BT100-648. This research has made use of the KMTNet system operated by the Korea Astronomy and Space Science Institute (KASI) and the data were obtained at three host sites of CTIO in Chile, SAAO in South Africa, and SSO in Australia. The Joan Or{\'o} Telescope (TJO) of the Montsec Astronomical Observatory (OAdM) is owned by the Catalan Government and operated by the Institute for Space Studies of Catalonia (IEEC). This publication also makes use of data products from NEOWISE, which is a project of the Jet Propulsion Laboratory/California Institute of Technology, funded by the Planetary Science Division of the National Aeronautics and Space Administration.
\end{acknowledgements}

\bibliographystyle{aa}
\bibliography{bibliograpahy_all}

\onecolumn
\begin{appendix} 
\section{Additional table}
\centering
\begin{longtable}{llllllll}
\caption{Observational and instrument details.}
\label{table:A1}\\
\hline\hline
            Date UT   & R. A. & Decl.  & $\Delta$     & $r_{h}$     & $\alpha$    & Site       & filter \\
                      & (h m) & ($^\circ$ ')  & (AU)  & (AU)  & ($^\circ$)  &            &        \\
\hline
\endfirsthead
\caption{continued.}\\
\hline\hline
            Date UT   & R. A. & Decl.  & $\Delta$     & $r_{h}$     & $\alpha$    & Site       & filter \\
                      & (h m) & ($^\circ$ ')  & (AU)  & (AU)  & ($^\circ$)  &            &        \\
\hline
\endhead
\hline
\endfoot

            May. 22, 2006 UT & 17 18 & -00 16 & 1.009 & 1.958 & 14.4 & SMO        & clear  \\
            May. 23, 2006 UT & 17 17 & +00 04 & 1.007 & 1.958 & 14.2 & SMO        & clear  \\
            May. 24, 2006 UT & 17 17 & +00 25 & 1.005 & 1.958 & 14.0 & SMO        & clear  \\
            May. 25, 2006 UT & 17 16 & +00 46 & 1.003 & 1.958 & 13.9 & SMO        & clear  \\
            May. 28, 2006 UT & 17 13 & +01 46 & 0.999 & 1.957 & 13.5 & SMO        & clear  \\
            May. 29, 2006 UT & 17 13 & +02 06 & 0.998 & 1.957 & 13.4 & SMO        & clear  \\
            May. 30, 2006 UT & 17 12 & +02 25 & 0.997 & 1.957 & 13.4 & SMO        & clear  \\
            Jun. 06, 2006 UT & 17 05 & +04 29 & 0.999 & 1.957 & 13.7 & SMO        & clear  \\
            Jun. 13, 2006 UT & 16 59 & +06 13 & 1.011 & 1.957 & 15.0 & Modra      & clear  \\
            Jun. 14, 2006 UT & 16 58 & +06 25 & 1.013 & 1.957 & 15.3 & Modra      & clear  \\
            Jun. 15, 2006 UT & 16 57 & +06 38 & 1.016 & 1.957 & 15.5 & Modra      & clear  \\
            Jun. 16, 2006 UT & 16 56 & +06 50 & 1.019 & 1.958 & 15.8 & SMO        & clear  \\
            Jun. 16, 2006 UT & 16 56 & +06 50 & 1.019 & 1.958 & 15.8 & Modra      & clear  \\
            Jun. 17, 2006 UT & 16 56 & +07 01 & 1.022 & 1.958 & 16.1 & Modra      & clear  \\
            Jun. 18, 2006 UT & 16 55 & +07 12 & 1.026 & 1.958 & 16.4 & SMO        & clear  \\
            Jun. 18, 2006 UT & 16 55 & +07 12 & 1.026 & 1.958 & 16.4 & Modra      & clear  \\
            Jun. 19, 2006 UT & 16 54 & +07 23 & 1.029 & 1.958 & 16.7 & SMO        & clear  \\
            Jun. 20, 2006 UT & 16 53 & +07 33 & 1.033 & 1.958 & 17.0 & Modra      & clear  \\
            Jun. 21, 2006 UT & 16 52 & +07 42 & 1.037 & 1.958 & 17.3 & SMO        & clear  \\
            Jun. 23, 2006 UT & 16 51 & +08 00 & 1.045 & 1.959 & 17.9 & Modra      & clear  \\
            Jun. 24, 2006 UT & 16 50 & +08 08 & 1.050 & 1.959 & 18.2 & Modra      & clear  \\
            Jun. 25, 2006 UT & 16 49 & +08 15 & 1.054 & 1.959 & 18.5 & Modra      & clear  \\
            Jul. 13, 2006 UT & 16 42 & +09 14 & 1.163 & 1.966 & 23.8 & Modra      & clear  \\
            Jul. 14, 2006 UT & 16 41 & +09 13 & 1.170 & 1.966 & 24.0 & Modra      & clear  \\
            Jul. 15, 2006 UT & 16 41 & +09 13 & 1.177 & 1.967 & 24.3 & Modra      & clear  \\
            Jul. 16, 2006 UT & 16 41 & +09 11 & 1.185 & 1.967 & 24.5 & Modra      & clear  \\
            Jul. 17, 2006 UT & 16 41 & +09 10 & 1.192 & 1.968 & 24.8 & Modra      & clear  \\
            Jul. 18, 2006 UT & 16 41 & +09 08 & 1.200 & 1.968 & 25.0 & Modra      & clear  \\
            Jul. 19, 2006 UT & 16 41 & +09 06 & 1.208 & 1.969 & 25.2 & Modra      & clear  \\
            Jul. 20, 2006 UT & 16 41 & +09 04 & 1.216 & 1.969 & 25.4 & Modra      & clear  \\
            Jul. 21, 2006 UT & 16 41 & +09 01 & 1.223 & 1.970 & 25.7 & Modra      & clear  \\
            Jul. 22, 2006 UT & 16 41 & +08 58 & 1.231 & 1.970 & 25.9 & Modra      & clear  \\
            Jul. 24, 2006 UT & 16 42 & +08 51 & 1.248 & 1.972 & 26.3 & Modra      & clear  \\
            Jun. 12, 2010 UT & 22 39 & +20 54 & 1.832 & 2.106 & 28.8 & WISE       & W1, W2, W3, and W4  \\
            Jan. 27, 2016 UT & 08 44 & -20 26 & 1.650 & 2.492 & 14.5 & KMTNet-SAAO       & R      \\
            Jan. 28, 2016 UT & 08 43 & -20 26 & 1.646 & 2.490 & 14.4 & KMTNet-SAAO       & R      \\
            Jan. 29, 2016 UT & 08 42 & -20 25 & 1.642 & 2.489 & 14.3 & KMTNet-SSO        & R      \\
            Jan. 29, 2016 UT & 08 42 & -20 25 & 1.642 & 2.489 & 14.3 & KMTNet-SAAO       & R      \\
            Feb. 03, 2016 UT & 08 37 & -20 15 & 1.626 & 2.481 & 14.0 & KMTNet-SSO        & R      \\
            Feb. 04, 2016 UT & 08 36 & -20 12 & 1.624 & 2.479 & 14.0 & KMTNet-SAAO       & R      \\
            Feb. 05, 2016 UT & 08 35 & -20 08 & 1.621 & 2.478 & 14.0 & KMTNet-SAAO       & R      \\
            Feb. 06, 2016 UT & 08 34 & -20 04 & 1.619 & 2.476 & 14.0 & KMTNet-SAAO       & R      \\
            Feb. 07, 2016 UT & 08 33 & -20 00 & 1.618 & 2.474 & 14.0 & KMTNet-SSO        & R      \\
            Feb. 07, 2016 UT & 08 33 & -20 00 & 1.617 & 2.474 & 14.0 & KMTNet-SAAO       & R      \\
            Feb. 08, 2016 UT & 08 32 & -19 55 & 1.616 & 2.473 & 14.0 & KMTNet-SSO        & R      \\
            Feb. 08, 2016 UT & 08 32 & -19 55 & 1.616 & 2.473 & 14.0 & KMTNet-SAAO       & R      \\
            Feb. 09, 2016 UT & 08 31 & -19 50 & 1.615 & 2.471 & 14.1 & KMTNet-SSO        & R      \\
            Feb. 10, 2016 UT & 08 30 & -19 45 & 1.613 & 2.470 & 14.1 & KMTNet-SSO        & R      \\
            Feb. 10, 2016 UT & 08 30 & -19 45 & 1.613 & 2.470 & 14.1 & KMTNet-SAAO       & R      \\
            Feb. 20, 2016 UT & 08 20 & -18 30 & 1.614 & 2.453 & 15.1 & KMTNet-SSO        & R      \\
            Feb. 22, 2016 UT & 08 19 & -18 16 & 1.616 & 2.451 & 15.3 & KMTNet-CTIO       & R      \\
            Feb. 22, 2016 UT & 08 19 & -18 12 & 1.615 & 2.451 & 15.2 & KMTNet-SSO        & R      \\
            Feb. 22, 2016 UT & 08 18 & -18 09 & 1.617 & 2.449 & 15.4 & KMTNet-SAAO       & R      \\
            Feb. 24, 2016 UT & 08 18 & -17 57 & 1.619 & 2.447 & 15.6 & KMTNet-CTIO       & R      \\
            Feb. 24, 2016 UT & 08 17 & -17 52 & 1.620 & 2.446 & 15.7 & KMTNet-SSO        & R      \\
            Feb. 24, 2016 UT & 08 17 & -17 49 & 1.621 & 2.446 & 15.8 & KMTNet-SAAO       & R      \\
            Feb. 28, 2016 UT & 08 15 & -17 16 & 1.629 & 2.441 & 16.3 & KMTNet-CTIO       & R      \\
            Feb. 28, 2016 UT & 08 15 & -17 11 & 1.630 & 2.440 & 16.4 & KMTNet-SSO        & R      \\
            Feb. 28, 2016 UT & 08 14 & -17 07 & 1.631 & 2.439 & 16.5 & KMTNet-SAAO       & R      \\
            Feb. 29, 2016 UT & 08 14 & -16 54 & 1.634 & 2.437 & 16.7 & KMTNet-CTIO       & R      \\
            Mar. 01, 2016 UT & 08 13 & -16 50 & 1.636 & 2.436 & 16.8 & KMTNet-SSO        & R      \\
            Mar. 01, 2016 UT & 08 13 & -16 45 & 1.637 & 2.436 & 16.9 & KMTNet-SAAO       & R      \\
            Mar. 03, 2016 UT & 08 13 & -16 32 & 1.641 & 2.434 & 17.1 & KMTNet-CTIO       & R      \\
            Mar. 03, 2016 UT & 08 12 & -16 27 & 1.642 & 2.433 & 17.2 & KMTNet-SSO        & R      \\
            Mar. 03, 2016 UT & 08 12 & -16 11 & 1.648 & 2.431 & 17.5 & KMTNet-SAAO       & R      \\
            Mar. 05, 2016 UT & 08 12 & -16 09 & 1.648 & 2.430 & 17.5 & KMTNet-CTIO       & R      \\
            Mar. 05, 2016 UT & 08 12 & -16 04 & 1.650 & 2.430 & 17.6 & KMTNet-SSO        & R      \\
            Mar. 07, 2016 UT & 08 11 & -15 45 & 1.656 & 2.427 & 17.9 & KMTNet-CTIO       & R      \\
            Mar. 07, 2016 UT & 08 11 & -15 36 & 1.660 & 2.426 & 18.1 & KMTNet-SAAO       & R      \\
            Mar. 11, 2016 UT & 08 10 & -14 57 & 1.674 & 2.420 & 18.8 & KMTNet-CTIO       & R      \\
            Mar. 15, 2016 UT & 08 09 & -14 07 & 1.695 & 2.413 & 19.6 & KMTNet-CTIO       & R      \\
            Mar. 15, 2016 UT & 08 09 & -13 57 & 1.700 & 2.412 & 19.7 & KMTNet-SAAO       & R      \\
            Mar. 22, 2016 UT & 08 09 & -12 35 & 1.740 & 2.400 & 21.0 & KMTNet-SSO        & R      \\
            Mar. 24, 2016 UT & 08 09 & -12 15 & 1.750 & 2.397 & 21.3 & KMTNet-CTIO       & R      \\
            Mar. 24, 2016 UT & 08 09 & -12 10 & 1.753 & 2.397 & 21.3 & KMTNet-SSO        & R      \\
            Mar. 24, 2016 UT & 08 09 & -12 05 & 1.756 & 2.396 & 21.4 & KMTNet-SAAO       & R      \\
            Mar. 26, 2016 UT & 08 09 & -11 51 & 1.764 & 2.394 & 21.7 & KMTNet-CTIO       & R      \\
            Mar. 26, 2016 UT & 08 09 & -11 46 & 1.766 & 2.393 & 21.8 & KMTNet-SSO        & R      \\
            Mar. 28, 2016 UT & 08 10 & -11 26 & 1.778 & 2.390 & 22.0 & KMTNet-CTIO       & R      \\
            Mar. 30, 2016 UT & 08 10 & -11 02 & 1.792 & 2.387 & 22.4 & KMTNet-CTIO       & R      \\
            Mar. 30, 2016 UT & 08 10 & -10 58 & 1.795 & 2.386 & 22.4 & KMTNet-SSO        & R      \\
            Apr. 03, 2016 UT & 08 12 & -10 06 & 1.829 & 2.378 & 23.1 & KMTNet-SAAO       & R      \\
            Apr. 04, 2016 UT & 08 13 & -09 48 & 1.841 & 2.375 & 23.3 & KMTNet-SSO        & R      \\
            Apr. 10, 2016 UT & 08 16 & -08 58 & 1.879 & 2.367 & 23.9 & KMTNet-CTIO       & R      \\
            Apr. 10, 2016 UT & 08 16 & -08 49 & 1.885 & 2.366 & 24.0 & KMTNet-SAAO       & R      \\
            Jun. 13, 2017 UT & 21 27 & +19 27 & 1.494 & 2.038 & 28.5 & LOAO       & R      \\
            Jun. 14, 2017 UT & 21 28 & +19 44 & 1.488 & 2.039 & 28.4 & BOAO       & R      \\
            Jun. 15, 2017 UT & 21 28 & +20 00 & 1.481 & 2.040 & 28.2 & BOAO       & R      \\
            Jun. 15, 2017 UT & 21 28 & +20 00 & 1.481 & 2.040 & 28.2 & LOAO       & R      \\
            Jun. 16, 2017 UT & 21 28 & +20 16 & 1.475 & 2.042 & 28.1 & BOAO       & R      \\
            Jun. 17, 2017 UT & 21 29 & +20 32 & 1.468 & 2.043 & 28.0 & BOAO       & R      \\
            Jun. 18, 2017 UT & 21 29 & +20 48 & 1.462 & 2.044 & 27.9 & BOAO       & R      \\
            Jun. 21, 2017 UT & 21 29 & +21 33 & 1.443 & 2.048 & 27.5 & La Sagra   & R      \\
            Jun. 22, 2017 UT & 21 29 & +21 48 & 1.437 & 2.050 & 27.3 & La Sagra   & R      \\
            Jun. 23, 2017 UT & 21 29 & +22 03 & 1.431 & 2.051 & 27.2 & La Sagra   & R      \\
            Jul. 10, 2017 UT & 21 26 & +25 29 & 1.341 & 2.076 & 24.3 & OAdM       & R      \\
            Jul. 12, 2017 UT & 21 25 & +25 47 & 1.333 & 2.079 & 23.9 & OAdM       & R      \\
            Jul. 15, 2017 UT & 21 24 & +26 11 & 1.320 & 2.084 & 23.4 & OAdM       & R      \\
            Jul. 17, 2017 UT & 21 23 & +26 25 & 1.313 & 2.087 & 23.0 & OAdM       & R      \\
            Jul. 24, 2017 UT & 21 18 & +26 59 & 1.290 & 2.098 & 21.6 & TUG        & R      \\
            Jul. 26, 2017 UT & 21 16 & +27 04 & 1.285 & 2.101 & 21.2 & TUG        & R      \\
            Aug. 14, 2017 UT & 21 00 & +26 15 & 1.265 & 2.132 & 18.4 & TUG        & R      \\
            Aug. 15, 2017 UT & 20 59 & +26 08 & 1.266 & 2.134 & 18.3 & TUG        & R      \\
            Aug. 24, 2017 UT & 20 52 & +24 41 & 1.280 & 2.149 & 17.9 & OAdM       & R      \\
            Aug. 24, 2017 UT & 20 52 & +24 41 & 1.280 & 2.149 & 17.9 & Mcd        & r      \\
            Aug. 25, 2017 UT & 20 51 & +24 29 & 1.282 & 2.151 & 17.9 & Mcd        & r      \\
            Aug. 26, 2017 UT & 20 50 & +24 17 & 1.285 & 2.153 & 17.9 & Mcd        & r      \\
            Aug. 27, 2017 UT & 20 50 & +24 05 & 1.287 & 2.155 & 17.9 & OAdM       & R      \\
            Aug. 27, 2017 UT & 20 50 & +24 05 & 1.287 & 2.155 & 17.9 & Mcd        & r      \\
            Aug. 28, 2017 UT & 20 49 & +23 53 & 1.290 & 2.156 & 17.9 & BlueEye600 & R      \\
            Aug. 28, 2017 UT & 20 49 & +23 53 & 1.290 & 2.156 & 17.9 & Mcd        & r      \\
            Aug. 29, 2017 UT & 20 49 & +23 40 & 1.293 & 2.158 & 18.0 & BlueEye600 & R      \\
            Aug. 29, 2017 UT & 20 49 & +23 40 & 1.293 & 2.158 & 18.0 & Mcd        & r      \\
            Aug. 30, 2017 UT & 20 48 & +23 26 & 1.297 & 2.160 & 18.0 & OAdM       & R      \\
            Sep. 09, 2017 UT & 20 44 & +21 01 & 1.340 & 2.178 & 18.8 & BOAO       & R      \\
            Oct. 02, 2017 UT & 20 47 & +14 57 & 1.505 & 2.219 & 22.1 & LOAO       & R      \\
            Oct. 25, 2017 UT & 21 04 & +09 51 & 1.744 & 2.262 & 24.7 & LOAO       & R      \\
\end{longtable}

\tablefoot{R. A. : Right ascension, Decl. : Declination, $\Delta$ : geocentric distance, $r_{h}$ : heliocentric distance, $\alpha$ : phase angle.}

\end{appendix}

\end{document}